# Nonlocal electrical detection of spin accumulation generated by Anomalous Hall effects in mesoscopic Ni$_{81}$Fe$_{19}$ films

Chuan Qin, Shuhan Chen,[*] Yunjiao Cai, Fatih Kandaz, and Yi Ji[†]

Department of Physics and Astronomy, University of Delaware, Newark, Delaware 19716

## Abstract

Spin accumulation generated by the anomalous Hall effects (AHE) in mesoscopic ferromagnetic Ni$_{81}$Fe$_{19}$ (permalloy or Py) films is detected electrically by a nonlocal method. The reciprocal phenomenon, inverse spin Hall effects (ISHE), can also be generated and detected all-electrically in the same structure. For accurate quantitative analysis, a series of nonlocal AHE/ISHE structures and supplementary structures are fabricated on each sample substrate to account for statistical variations and to accurately determine all essential physical parameters *in-situ*. By exploring Py thicknesses of 4 nm, 8 nm, and 12 nm, the Py spin diffusion length $\lambda_{Py}$ is found to be much shorter than the film thicknesses. The product of $\lambda_{Py}$ and the Py spin Hall angle $\alpha_{SH}$ is determined to be independent of thickness and resistivity: $\alpha_{SH}\lambda_{Py}$= (0.066 ± 0.009) nm at 5 K and (0.041 ± 0.010) nm at 295 K. These values are comparable to those obtained from mesoscopic Pt films.

---

[*] Present address: Western Digital Corporation, 44100 Osgood Rd, Fremont CA 94539
[†] Email: yji@udel.edu

**I, Introduction**

A pure spin current, which is a flow of spin angular momenta without a net charge current, provides important functionalities for spintronics. Recently, spin Hall effect (SHE) and inverse spin Hall effects (ISHE) have been explored extensively for the conversion between charge current and pure spin current. [1-13] SHE converts longitudinal charge current into transverse spin current. The reciprocal process, inverse spin Hall effect (ISHE), converts spin current into charge current. The SHE/ISHE originates from the strong spin-orbit coupling and was initially studied in heavy nonmagnetic metals such as Pt.[6-13] Later work shows that ISHE also exists in ferromagnetic metal such as $Ni_{81}Fe_{19}$ alloy (permalloy or Py),[14-16] which contains lighter elements than Pt. It is an intriguing prospect, because Py is less expensive than Pt and is a commonly used material in spintronics. This also implies that a transverse spin current would coexist with the transverse charge current produced by the anomalous Hall Effect (AHE), which is the reciprocal of the ISHE in a ferromagnet. A direct measurement of spin current or spin accumulation from the AHE is important to spintronics, because the interplay between the spin current from AHE and the anisotropic magnetoresistance in a ferromagnetic metal is predicted to lead to versatile spin transfer switching.[17] The spin-orbit effects that give rise to AHE can also induce additional torques in the spin dynamics driven by short magnetic pulses.[18]

However, the ferromagnetic nature of Py complicates experimental efforts of probing the spin current that accompanies AHE. Previous relevant work was conducted in the context of ISHE and has used bilayers of Py and ferromagnetic insulator yttrium iron garnet (YIG). A pure spin current from YIG is produced by a temperature gradient via spin Seebeck effect[14, 16] or by microwave excitation via spin pumping.[15] Because of the ISHE, a charge voltage is generated as the pure spin current propagates through Py. The choice of ferromagnetic insulator avoids entanglements of the magneto-resistive effects from otherwise two ferromagnetic metals. However, direct detection of spin accumulation or spin current induced by AHE in a ferromagnet is still lacking.

In this work, a nonlocal method is used to directly measure the spin current generated from mesoscopic Py films by AHE. In the same structure, The ISHE in Py is generated and detected *electrically*, complementing previously used spin-Seebeck and spin-pumping methods. With the alternating current (AC) modulation method, the AHE/ISHE signals are extracted from the linear response of the nonlocal voltage difference between two polarities of large magnetic fields. Therefore, the signals are well separated from anomalous Nernst effects, anisotropic magnetoresistance, or regular nonlocal spin signals.

The strength of the SHE/ISHE is often described by a spin Hall angle $\alpha_{SH} = \sigma_{SH}/\sigma = \sigma_{SH}\rho$, where $\sigma_{SH}$ is the spin Hall conductivity, and $\sigma$ and $\rho$ are the electrical conductivity and resistivity, respectively. Equally important is the spin diffusion length $\lambda$ of the SHE/ISHE material. For a thin film that is substantially thicker than $\lambda$, the SHE/ISHE can be enhanced by either increasing $\alpha_{SH}$ or increasing $\lambda$. Overestimating one leads to underestimating the other. Furthermore, the spin diffusion length is unlikely to be a material constant, because it varies with the dimension and the resistivity of the material. In this work, we use the product of spin Hall angle and the spin diffusion length $\alpha_{SH}\lambda$, or equivalently $\sigma_{SH}\rho\lambda$, as a figure of merit to quantify the AHE/ISHE in Py.

Accurate quantitative analysis hinges on the accurate determination of all relevant physical quantities, as well as proper treatment of statistical variations between devices.

To this end, several (6 - 8) nonlocal AHE/ISHE structures are fabricated on each sample substrate to account for the statistical variations between structures. Supplementary structures (~ 20) are fabricated on the same substrate to provide accurate measurements of the *in-situ* values of Py and Cu resistivity $\rho_{Py}$ and $\rho_{Cu}$, spin diffusion length $\lambda_{Cu}$ of Cu, and spin polarizations *P* and resistance $R_i$ of Py/Cu interfaces. The resistivity of mesoscopic Py films is substantially higher than that of extended films, but decreases as the film thickness increases. By exploring different Py thicknesses, $\lambda_{Py}$ is found to be ≤ 1.0 nm. The value of $\sigma_{SH}\rho_{Py}\lambda_{Py}$ is independent of thickness and resistivity, and comparable to the $\sigma_{SH}\rho_{Pt}\lambda_{Pt}$ obtained previously for mesoscopic Pt films.

**II, Sample Preparation**

The nonlocal AHE/ISHE structures along with two types of supplementary structures are fabricated simultaneously on a single substrate by using shadow mask techniques.[19] Two-layer e-beam resists, PMMA (polymethyl methacrylate) on the top and PMGI (polydimethylglutarimide) at the bottom, are coated on the silicon substrate covered with 200 nm $Si_3N_4$. Mesoscopic suspended shadow masks are formed in the resist layers after electron beam (e-beam) lithography, because large undercut develops in the PMGI layer. The shadow mask for the AHE/ISHE structures is illustrated in Figure 1 (a). On the same substrate, additional shadow masks are formed for supplementary structures, which are the nonlocal spin valves (NLSV)[20-25] and the Py resistivity measurement structures.

Deposition of various materials through the shadow masks is carried out from different angles to form the structures without breaking vacuum. First, Py is evaporated from opposite oblique angles to form two Py pads, designated as Py1 and Py2, as shown in Figure 1 (a) - (d). Subsequently 3 nm $AlO_x$ and 110 nm Cu are deposited along the substrate normal direction. Therefore, all interfaces and materials are formed in high vacuum to ensure efficient spin transport. The scanning electron microscope (SEM) images of finished AHE/ISHE structures, NLSVs, and Py resistivity measurement structures are shown in Figure 1 (b), (c), and (d), respectively. More details of shadow evaporation method can be found elsewhere.[26-29]

For the AHE/ISHE structure in Figure 1 (b), the Py1 is the spin injector for the ISHE measurement or the spin detector for the AHE measurement, and the thickness remains 12 nm for all samples. The Py2 electrode is the anomalous/spin Hall material in which the AHE or ISHE is generated, and thicknesses of 4 nm, 8 nm, and 12 nm are used on different sample substrates. The Cu channel is used to transport a pure spin current between Py1 and Py2. The widths for Py1, Py2, and Cu are ~250 nm, ~230 nm, and ~80 nm, respectively.

The NLSV, shown in Figure 1 (c), consists of two Py1 electrodes (an injector and a detector) and a Cu channel, and is used to determine the spin polarization $P$ at the Py1/AlO$_x$/Cu interfaces and the spin diffusion length $\lambda_{Cu}$ and resistivity $\rho_{Cu}$ of Cu channel. The Py resistivity measurement structure, shown in Figure 1 (d), is a mesoscopic Py2 stripe with four electrical probes. It is used to determine the resistivity $\rho_{Py}$ of the mesoscopic Py2 film. The $P$, $\lambda_{Cu}$, $\rho_{Cu}$, and $\rho_{Py}$ determined from the supplementary structures are the same as those in the AHE/ISHE structures, because all structures undergo identical fabrication procedures. The lateral dimensions of all structures are characterized by SEM for quantitative analysis.

The 3 nm AlO$_x$ layers in all structures are directly evaporated from AlO$_x$ pellets by electron beam, and the typical resistance for a 100 nm × 100 nm junction is between 5 and 20 Ω. Therefore, it is not a uniform tunnel barrier, which should have much higher resistance. However, it has been demonstrated that the low-resistance AlO$_x$ interfaces are more effective than transparent ohmic interfaces in preventing absorption of spin current into the magnetic electrodes of the NLSV.[27, 30-32] Therefore, a higher spin accumulation can be maintained in the Cu channel, leading to large spin signals in NLSV. In addition, an electric current that flows through the Py2 electrode and generates AHE can be undesirably shunted by the conductive Cu, but the finite resistance of Cu/AlO$_x$/Py2 interfaces reduces this shunting effect.

**III, Measurements**

Measurements are performed in a pulse-tube variable temperature cryostat at 5 K and 295 K. We describe the measurements using the 5 K results for the sample with 8 nm Py2. Figure 2 (a) and (b) illustrate the ISHE and AHE measurements, respectively. The measurement configurations are shown in the insets. For ISHE, an injection current is directed between the Py1 (I+) and the upper end of the Cu channel (I-). The nonlocal voltage is detected between the two ends of the Py2 stripe, with V+ on the left and V- on the right. For AHE, the current flows through the Py2 stripe, with I+ on the right and I- on the left. The nonlocal voltage is measured between the Py1 (V+) and the upper end of Cu channel (V-). An alternating current (*a.c.*) of $I_e$ = 0.2 mA with a frequency of 346.5 Hz is applied, and the nonlocal *a.c.* voltage $V_{nl}$ is detected by lock-in method. The nonlocal resistance, $R_s = V_{nl}/I_e$, is recorded as a function of the magnetic field $B_x$, which is applied parallel to the Cu channel.

There are two major features of the $R_s$ versus $B_x$ curve in Figure 2 (a). One is the double-dips at the intermediate fields, which results from conventional nonlocal spin signals. The other feature is the difference of $\Delta R_s = 0.71$ mΩ between positive and negative high fields, which results from the ISHE. The nonlocal spin signal is symmetrical in the sense that the $R_s$ values at large positive and negative fields are equal, as illustrated in the top portion of Figure 2 (c). The SHE/ISHE signals are asymmetrical with different $R_s$ values at large positive and negative fields, as shown in the middle portion of Figure 2 (c). The overall $R_s$ versus $B_x$ curve is a superposition of the symmetrical nonlocal spin signal and the asymmetrical ISHE signal, as illustrated in the bottom portion of Figure 2 (c). Therefore, the ISHE signal can be clearly separated from the conventional nonlocal spin signal.

It is useful to explain why nonlocal spin signals can be detected between the two ends of Py2 electrode, considering that the more standard practice is to measure it between the Py2 detector and the Cu channel. One can imagine that a nonlocal spin signal can be measured in the more standard way between the left end of Py2 and the Cu channel, referring to the inset of Figure 2 (a). Similarly, another nonlocal spin signal can be measured between the right end of Py2 and the Cu channel. The two signals should be the same if the Cu/AlO$_x$/Py2 interface is a uniform tunnel barrier with large resistance. However, the directly evaporated AlO$_x$ layer is less than ideally uniform and the resistance (typically 5 – 20 Ω for a 100 nm × 100 nm junction) is much lower than that of tunnel barriers. The signals measured from the left and the right are strongly affected by the interface conditions on the two sides of the junction, and typically have different values. Therefore, the signal measured between two ends of Py2, as in this work, is equivalent to a subtraction of the two different signals from left and right. More detailed analysis can be found in our previous work by Chen *et al.*[33]

The nonlocal spin signals are symmetrical because parallel states between the spin injector and spin detector are equivalently reached at large positive and negative fields. For SHE, however, the spin accumulation generated by the charge current is unaffected by a reversal of magnetic field, but the magnetization of the spin detector can be switched by the field. This apparent asymmetry of the system leads to the different (asymmetrical) $R_s$ values at large positive and negative fields, which is the signature for SHE/ISHE in nonlocal structures.[7, 28, 34, 35]

The $R_s$ reaches negative values around the dips of the curve in Figure 2 (a). This is routinely observed in a nonlocal measurement, which should have zero charge voltage background in an ideal situation. In experiments, the background voltage (or baseline) is often close to but not exactly zero.[27, 36, 37] Any spin-related signal change, either nonlocal spin signal or SHE/ISHE signal, is likely to swing the measured voltage between positive and negative $R_s$ values. Such change of signal sign is an indication of clean nonlocal measurements rather than artifacts.

In previous work on ISHE in Py with Seebeck method, anomalous Nernst effects can be present and have to be explicitly separated or ruled out.[14, 16, 38] In this nonlocal method, the detected nonlocal voltage is locked to the base frequency of the sinusoidal excitation currents. Thermal effects are excluded, because thermal effects are proportional to the square of excitation current and therefore related to the voltage response at second harmonics.

Figure 2 (b) shows the $R_s$ versus $B_x$ curve in the AHE measurement. Because of the Onsager reciprocal relations, the curve yields the same asymmetrical difference of $\Delta R_s = 0.71$ m$\Omega$ and the same magnitude of symmetrical nonlocal signals as compared to Figure 2 (a). The nonlocal signal is inverted (peaks instead of dips) because the electrical polarities on Py2 are inverted between the AHE and ISHE configurations. Following previously used conventions,[7, 34, 35] the AHE/ISHE signal is defined as $\Delta R_{AHE} = \frac{\Delta R_S}{2} = 0.305$ m$\Omega$. For each Py2 thickness, 6 - 8 AHE/ISHE devices are measured for quantitative analysis.

At positive or negative large fields, the magnetization of Py2 electrode is oriented to opposite directions ($\pm x$). The transverse charge voltages between the top and bottom surfaces of Py2, induced by the AHE, have opposite signs for opposite magnetizations. However, the spin currents in the $z$ direction from AHE should be the same. The majority and minority spins move in opposite directions but contribute positively to the transverse spin current. Reversed magnetization switches the roles of majority and minority spins, but will not alter the net spin current or spin accumulation. Therefore, the treatment of spin accumulation from AHE is identical to that of SHE, and we use the term AHE and SHE interchangeably throughout this paper.

The resistance $R_i$ of the Cu/AlO$_x$/Py2 interface is an important quantity to estimate the spin current through the interface and the current shunting effect by the Cu, and it can be measured directly in each AHE/ISHE structure. A current is applied between the right side of Py2 strip and the upper end of Cu channel, referring to Figure 1 (b), and a voltage is detected between the left sides of Py2 and Py1 electrodes. Depending on the size of Cu/AlO$_x$/Py2 interface, $R_i$ varies between 2 $\Omega$ and 130 $\Omega$. The values of $R_i$ for the substrate with 8 nm Py2 are summarized in Table I.

A set (8 – 12) of supplementary NLSVs are used to determine spin polarization $P$ of the Py1/AlO$_x$/Cu interface as well as $\lambda_{Cu}$, and $\rho_{Cu}$. Spin signals are measured as a function of center-to-center distance $L'$ between two Py1 electrodes. Figure 3 (a) shows a $R_s$ versus $B_y$ curve of a NLSV at 5 K and the measurement configuration is illustrated in the inset. The high $R_s$ value is associated with the parallel state between the injector and detector, the low $R_s$ is associated with the anti-parallel state, and the difference $\Delta R_s$ is the NLSV spin signals. By fitting $\Delta R_s$ versus $L'$ with the equation $\Delta R_s = (P^2 \rho_{Cu} \lambda_{Cu} / A_{Cu}') exp(-L'/\lambda_{Cu})$, we are able to extract $P$ and $\lambda_{Cu}$. The cross-sectional area of the NLSV Cu channel is $A_{Cu}' = t_{Cu} w_{Cu}'$, where $t_{Cu} = 110$nm is the thickness and $w_{Cu}'$ is the width. The values of $L'$ and $w_{Cu}'$ are measured by SEM for each device. The value of $\rho_{Cu}$ is determined by sending a current through the Cu channel and detecting the voltage between two Py electrodes. An average $\rho_{cu}$ is determined for devices on each sample substrate and the values are in the range of 1.8 – 2.6 µΩ•cm at 5 K. These values are reasonably small because of the large thickness of Cu. Figure 3 (b) shows the $\Delta R_s$ versus $L'$ plot, and the fitting yields $P$ = 16.5% and $\lambda_{Cu}$ = 900 nm at 5 K for the sample substrate with 8 nm Py2.

Four probe measurements are performed on 8 – 12 Py resistivity structures to determine the average $\rho_{Py}$ of Py2 stripes, which have the same width as the Py2 in AHE/ISHE structures. The measured average $\rho_{Py}$ as a function of the Py2 thickness $t_{Py}$ is plotted in the inset of Figure 4 (a) for 5 K and (b) for 295 K. The values of $\rho_{Py}$ are between 150 µΩ•cm and 470 µΩ•cm, and decrease with an increasing $t_{Py}$. These values are 4 – 8 times larger than that of thick extended films. Reduction of either thickness or width leads to an increase of resistivity because of surface and edge defects. Therefore, it is important to measure resistivity on films that bear the same thickness and width as the AHE/ISHE structures.

## IV, Results and Analysis

In our previous work,[35] we developed an approach to quantitatively analyze the SHE/ISHE of mesoscopic Pt thin films in nonlocal structures. Spin accumulation in Cu channels and Pt thin films can be solved using one-dimensional spin diffusion equations with proper boundary conditions. Also, the spin current across the Cu/AlO$_x$/Pt interface and charge current/voltage shunting near the interface can be well quantified by the resistance $R_i$ of the Cu/AlO$_x$/Pt interface.

Using the same method, the AHE/ISHE signal in Py can be expressed as:

$$\Delta R_s = 2\Delta R_{AHE} = \frac{\alpha'_{SH} \gamma \chi \rho_{Py} P}{2 w_{Py}} exp\left(-\frac{L}{\lambda_{Cu}}\right) \quad (1)$$

with the definition of the apparent spin Hall angle $\alpha'_{SH}$:

$$\alpha'_{SH}(t_{Py}) = 2\alpha_{SH}\left(\frac{\lambda_{Py}}{t_{Py}}\right)\left(\frac{exp(t_{Py}/\lambda_{Py})-1}{exp(t_{Py}/\lambda_{Py})+1}\right) \quad (2).$$

Here the spin absorption coefficient $\gamma = 2R_{sCu}/(R_{sCu} + R_i)$ describes the amount of spin current across the Cu/AlO$_x$/Py2 interface, and $L$ is the center-to-center distance between the Py1/AlO$_x$/Cu and Cu/AlO$_x$/Py2 junctions. The $R_{sCu} = \rho_{Cu}\lambda_{Cu}/A_{Cu}$ is the Cu spin resistance with $A_{Cu} = t_{Cu}w_{Cu}$ being the Cu cross-sectional area. The factor $\chi = 4R_i/(4R_i + R_{Py})$ describes the shunting effect to the ISHE voltage or the AHE-inducing current by the highly conductive Cu through the Cu/AlO$_x$/Py2 interface.[35] The $R_{Py}$ is defined as $R_{Py} = \rho_{Py}w_I/(w_{Py}t_{Py})$, where $w_I$ is the width of Cu above the Cu/AlO$_x$/Py2 interface and $w_{Py}$ is the overall width of the Py2 stripe. The values of $w_{Cu}, w_{Py}, w_I$ and $L$ are carefully measured by SEM for each AHE/ISHE device.

It is useful to consider two limiting cases for Eq. 2. When $t_{Py} \ll \lambda_{Py}$, $\alpha'_{SH} = \alpha_{SH}$ and the apparent spin Hall angle is a constant and equals to spin Hall angle. When $t_{Py} \gg$ $\lambda_{Py}$, $\alpha'_{SH}(t_{Py}) = 2\alpha_{SH}\left(\frac{\lambda_{Py}}{t_{Py}}\right) = \frac{2\sigma_{SH}\rho_{Py}\lambda_{Py}}{t_{Py}}$ and $\alpha'_{SH}$ is inverse proportional to $t_{Py}$. Even in the case of varying $\rho_{Py}$ with $t_{Py}$, the value of $\rho_{Py}\lambda_{Py}$ would still be a constant under the assumption of Elliot-Yafet spin relaxation mechanism[39, 40] with a fixed spin relaxation rate. Without precise knowledge of $\lambda_{Py}$, we first calculate the $\alpha'_{SH}$ from measured $\Delta R_{AHE}$ using Eq. 1. The obtained $\alpha'_{SH}$ at 5 K for devices on the sample substrate with 8 nm Py2 are shown in Table I. The average $\alpha'_{SH}$ as a function of $t_{Py}$ are shown in Figure 4 (a) for 5 K and Figure 4 (b) for 295 K. For both temperatures, a drastic decay is observed and indicates a short spin diffusion length ($\lambda_{Py} \ll t_{Py}$). The data is fitted well (solid lines) by a $\sim 1/t_{Py}$ dependence. We also attempted to fit the $\alpha'_{SH}$ versus $t_{Py}$ data by using Eq. 2 with an assumed $\lambda_{Py}$ and a free fitting parameter $\alpha_{SH}$. When using $\lambda_{Py}$ values lower than 1.0 nm, we could obtain equally good fits. When using $\lambda_{Py} = 1.2$ nm or higher, the fitted curves obviously deviate from experimental data. Therefore, we conclude that $\lambda_{Py}$ is no more than 1.0 nm and much shorter than the film thickness (4 -12 nm). The short $\lambda_{Py}$ is also consistent with high resistivity $\rho_{Py}$ measured in mesoscopic Py films, because the Elliott-Yafet model implies that $\rho\lambda$ is a constant. Note that the fabrication does not involve etching that may degrade the quality of Py films.

In the short $\lambda_{Py}$ limit: the relation between $\alpha'_{SH}$ and $t_{Py}$ can be rewritten as $\alpha_{SH}\lambda_{Py} = \sigma_{SH}\rho_{Py}\lambda_{Py} = \alpha'_{SH}t_{Py}/2$, which allows us to obtain $\alpha_{SH}\lambda_{Py}$ or $\sigma_{SH}\rho_{Py}\lambda_{Py}$ for each $t_{Py}$ from experimental results, without any assumption of fixed values of $\sigma_{SH}$, $\rho_{Py}$, $\lambda_{Py}$, or their products. The obtained average values of $\alpha_{SH}\lambda_{Py}$ are plotted as a function of $t_{Py}$ in Figure 4 (c), giving a constant value $\alpha_{SH}\lambda_{Py} = (0.066 \pm 0.009)$ nm at 5 K and $\alpha_{SH}\lambda_{Py} = (0.041 \pm 0.010)$ nm at 295 K. If $\lambda_{Py} = 1.0$ nm, the spin Hall angle $\alpha_{SH}$ would be 0.066 at 5 K and 0.041 at 295 K. For shorter $\lambda_{Py}$, the $\alpha_{SH}$ would be higher accordingly. If we continue to assume a constant $\rho_{Py}\lambda_{Py}$, the constant values of $\sigma_{SH}\rho_{Py}\lambda_{Py}$ imply that $\sigma_{SH}$ is independent of thickness and resistivity. This is consistent with the intrinsic or side jump mechanism in spin Hall or anomalous Hall effects.[41] For skew scattering mechanism, the $\sigma_{SH}$ is inversely proportional to electrical resistivity. The intrinsic or side jump mechanism is expected to dominate in the moderately dirty conductors with relatively high resistivity.

Previously, we explored and analyzed the SHE/ISHE in mesoscopic Pt thin films at 5 K using the same method.[35] In Figure 5, we compare the values of $\alpha_{SH}\lambda$ versus film thickness at 5 K for Pt and Py films. At lower thickness (4 nm), the ratio of two values is $(\alpha_{SH}\lambda)_{Py}/(\alpha_{SH}\lambda)_{Pt} \approx 0.47$; At higher thickness (12 nm), the ratio is 0.93, indicating that the spin current from the AHE in mesoscopic Py films is comparable to that from the SHE in mesoscopic Pt films.

The value of $\alpha_{SH}\lambda_{Py} = 0.041$ nm at 295 K can be compared to results obtained on Py/YIG bilayer structures. By using spin Seebeck effects and ISHE, Miao *et al.*[14] reported $\alpha_{SH} = 0.005$ and $\lambda_{Py} = 2.5$ nm, yielding $\alpha_{SH}\lambda_{Py} = 0.0125$ nm. Wang *et al.*[15] used the spin pumping measurements and obtained $\alpha_{SH} = 0.02$ and $\lambda_{Py} = 1.7$ nm, thereby giving $\alpha_{SH}\lambda_{Py} = 0.034$ nm. These values are lower than ours but are on the same order of magnitude.

**V, Conclusion**

In summary, large spin accumulation caused by Anamalous Hall effect has been detected electrically using a nonlocal method in mesoscopic NiFe (Py) thin films. Its reciprocal effects, the inverse spin Hall effects, are also generated and detected. A systematic approach is used to quantify the effects and obtain the product of spin Hall angle and the spin diffusion length: $\alpha_{SH}\lambda_{Py}$= (0.066 ± 0.009) nm at 5 K and (0.041 ± 0.010) nm at 295 K. These values are independent of film thickness and resistivity, and are comparable to that of mesoscopic Pt films.

**Acknowledgements**

Fatih Kandaz acknowledges support by Republic of Turkey Ministry of National Education.

**Table I**

| Device | Junction Size | $\Delta R_s$ | L | $R_i$ | $\chi$ | $\gamma$ | $\alpha_{SH}'$ |
|---|---|---|---|---|---|---|---|
| | | mΩ | nm | Ω | | | |
| 3-1 | 50nm | 0.44 | 429 | 43.1 | 0.74 | 0.076 | 0.017 |
| 2-1 | 50nm | 0.30 | 431 | 74.2 | 0.84 | 0.045 | 0.017 |
| 1-6 | 100nm | 0.52 | 444 | 21.5 | 0.58 | 0.14 | 0.014 |
| 5-7 | 150nm | 0.57 | 448 | 3.6 | 0.16 | 0.57 | 0.015 |
| 6-7 | 150nm | 0.61 | 444 | 8.8 | 0.32 | 0.27 | 0.017 |
| 7-2 | 200nm | 0.71 | 450 | 1.5 | 0.065 | 0.98 | 0.025 |

**Table Caption**

Table I: Measured AHE/ISHE signals $\Delta R_s$ at 5 K, various parameters, and the obtained apparent spin Hall angle $\alpha_{SH}'$ for 6 AHE/ISHE devices on the substrate with 8 nm thick Py2. The junction size refers to the length of the Cu channel right above the Py2/AlO$_x$/Cu junction.

**Figure Captions**

**Figure 1.** (a) Angle evaporation through a mesoscopic suspended shadow mask designed for the AHE/ISHE structures. SEM images for (b) an AHE/ISHE structure, (c) a NLSV and (d) a Py resistivity measurement structure. All structures are fabricated on the same substrate.

**Figure 2.** The $R_s$ versus $B_x$ curves at 5 K of (a) an ISHE measurement and (b) an AHE measurement from the same AHE/ISHE structure with 8 nm thick Py2. The insets of (a) and (b) show the measurement configurations. The magnetic field is applied parallel to Cu channel ($\pm x$ direction). (c) Illustration of symmetrical nonlocal spin signal (top), asymmetrical SHE/ISHE signal (middle), and their superposition (bottom).

**Figure 3.** (a) The $R_s$ versus $B_y$ curve at 5 K for a NLSV with magnetic field $B_y$ applied along the $\pm y$ direction (shown in the inset). The blue arrows indicate the magnetization states of the injector and the detector. (b) The $\Delta R_s$ versus $L'$ and a fit (solid red line) for NLSVs on the substrate with 8 nm thick Py2.

**Figure 4.** The apparent spin Hall $\alpha'_{SH}$ as a function of Py thickness (a) at 5K and (b) at 295K and fits (solid lines). The insets show the Py resistivity $\rho_{Py}$ as a function of Py thickness. (c) The obtained values of $\alpha_{SH} \lambda_{Py}$ as a function of $t_{Py}$ at 5 K (red squares) and 295 K (black dots). The dash-dot line is a guidance for the eyes.

**Figure 5.** A comparison of $\alpha_{SH}\lambda$ between mesoscopic Py and Pt films as a function of film thickness at 5 K.

**Figures**

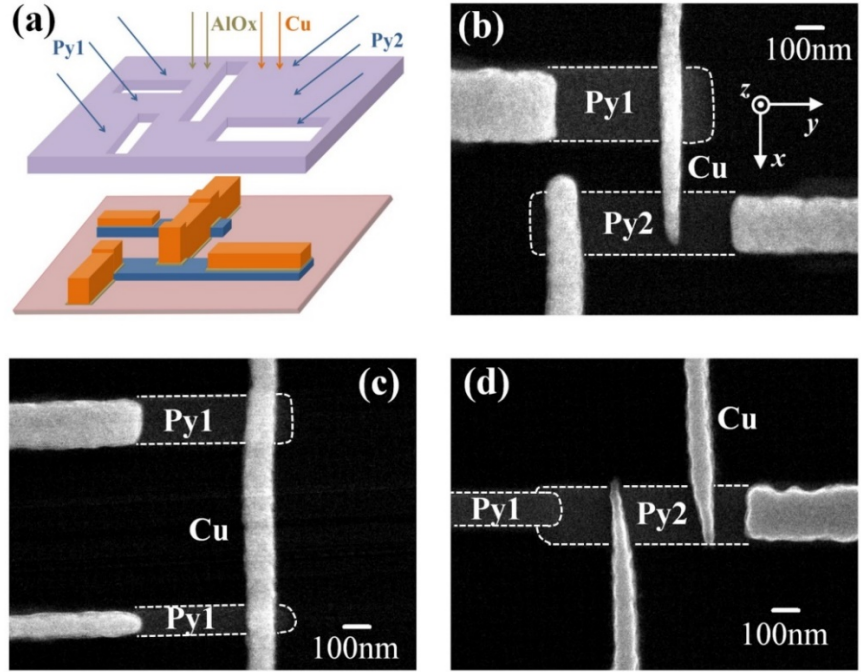

**Figure 1**

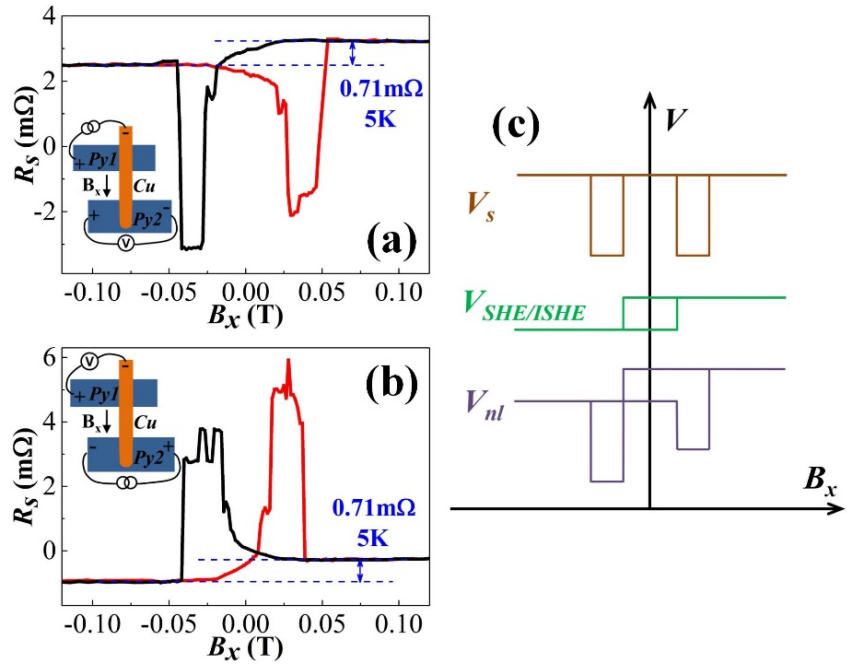

**Figure 2**

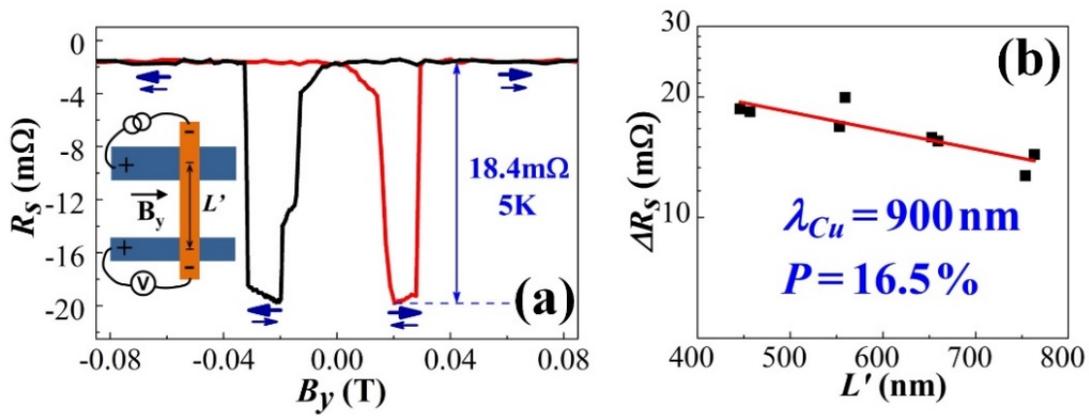

Figure 3

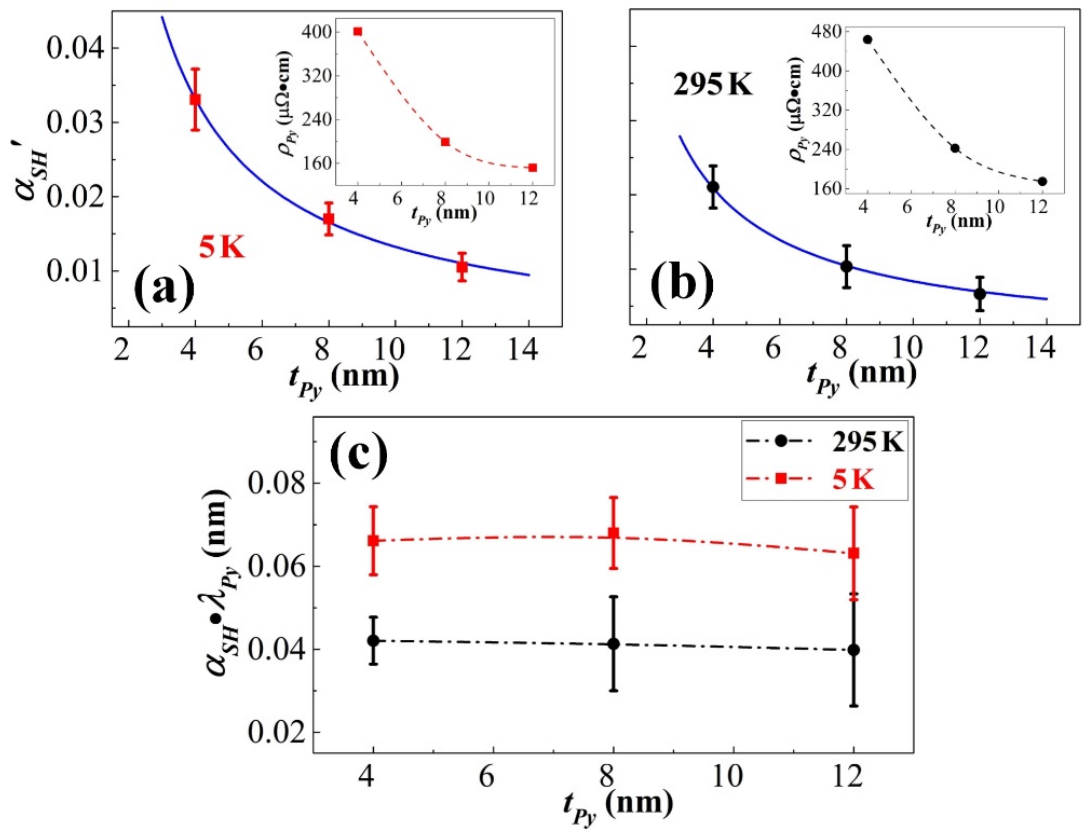

Figure 4

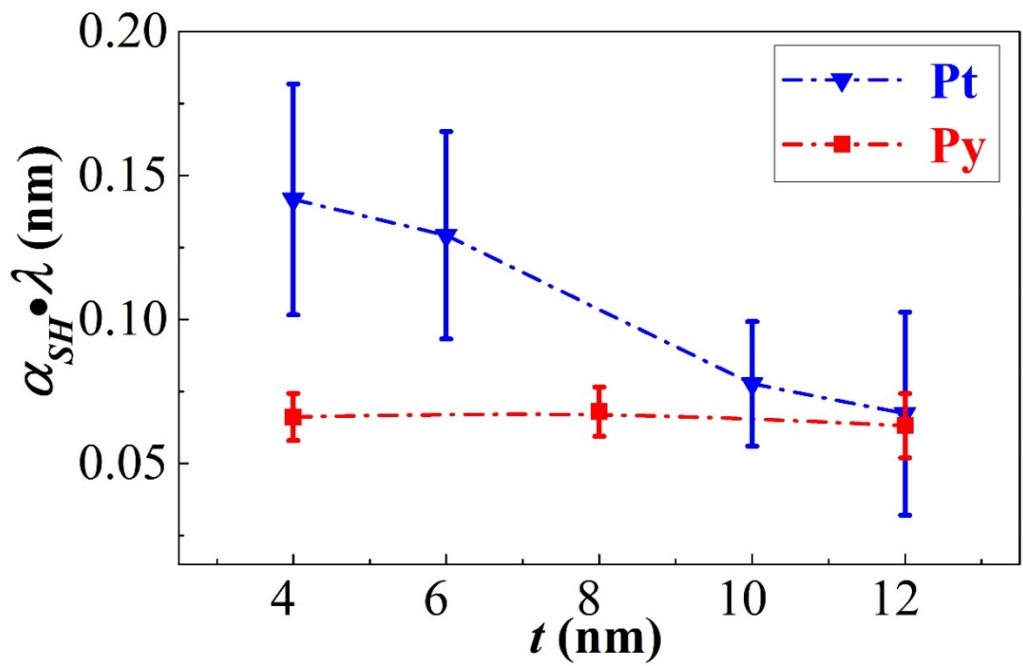

**Figure 5**